\documentclass{JHEP3} % 10pt is ignored!
\usepackage{epsfig}
\epsfclipon

\newcommand{\nco}{\newcommand}
\nco{\beq}{\begin{equation}} \nco{\eeq}{\end{equation}}
\nco{\beqa}{\begin{eqnarray}} \nco{\eeqa}{\end{eqnarray}}
\nco{\lra}{\leftrightarrow}
\def\sfrac#1#2{{\textstyle{#1\over #2}}}

\nco{\sss}{\scriptscriptstyle} \nco{\dphi}{\varphi}
\nco{\lsim}{\mbox{\raisebox{-.6ex}{~$\stackrel{<}{\sim}$~}}}
\nco{\gsim}{\mbox{\raisebox{-.6ex}{~$\stackrel{>}{\sim}$~}}}

\def\pref#1{(\ref{#1})}

\def\Hp{H_+}
\def\hp{h_+}
\def\Hm{H_-}
\def\hm{h_-}

\def\exd{{\rm d}}

\title{\Large Decoupling, Trans-Planckia and Inflation\footnote{Invited talk
given by C.B. at the Davis Meeting on Cosmic Inflation, March
2003.}}

\author{C.P.\ Burgess, James M.~Cline, F. Lemieux\\
Physics Department, McGill University,
3600 University Street, Montr\'eal, Qu\'ebec, Canada H3A 2T8\\
E-mail: \email{cliff@physics.mcgill.ca},
\email{jcline@physics.mcgill.ca},
\email{lemieux@physics.mcgill.ca}}
\author{and R.~ Holman\\ Physics Department, Carnegie Mellon
University, Pittsburgh PA 15213\\
E-mail: \email{rh4a@andrew.cmu.edu}}

\preprint{McGill-03/13}


\abstract{We survey recent calculations probing what constraints
decoupling can put on the influence of very-high-energy physics on
the predictions of inflation for the cosmic microwave background.
Using garden-variety hybrid inflation models we identify two ways
in which higher-energy physics can intrude into inflationary
predictions. 1.\ Non-adiabatic physics up to 30 $e$-foldings
before horizon exit can have observable consequences for the CMB,
including the introduction of features in the fluctuation spectrum
at specific multipoles and a general suppression of power at large
scales (a prediction which was made {\it before} the recent
release of WMAP results). Our comparison of simple models with the
data marginally improves the goodness of fit compared to the
standard concordance cosmology, but only at the 1.5-sigma level.
2.\ Adiabatic physics can also affect inflationary predictions
through virtual loops of very-heavy particles, but these can only
be distinguished from lower-energy effects within the context of
specific models. We believe our conclusions should apply equally
well to trans-Planckian physics provided only that this physics
satisfies decoupling, such as string theory appears to do.
(Non-decoupling trans-Planckian proposals must explain why
meaningful theoretical predictions at low energies are possible at
all.)}

\begin{document}

\section{Introduction and Discussion \label{section:intro}}
The brave new world of precise measurements of the cosmic
microwave background (CMB) \cite{BNW} has motivated many studies
of what the theoretical implications of these measurements might
be. In particular, the observations agree very well --- apart from
the controversial evidence for a measurement of nonzero $\exd
n/\exd\ln k$ in the WMAP results \cite{Spergeletal} --- with the
predictions of generic inflationary models, and their precision is
beginning to differentiate amongst the various models which have
been proposed \cite{WMAPInflation}. This may well be our first
direct observation of the physics of energies which are extremely
high compared to those to which we have experimental access
elsewhere.

Any meaningful quantitative comparison between models and
observations requires a clear understanding of the theoretical
uncertainties which are involved, and in this context a recent
controversy has emerged about whether the successful inflationary
predictions are subject to uncontrollable theoretical errors. The
controversy was initiated by various calculations
\cite{tp1,tp2,tp3} claiming that observable effects are possible
for the CMB spectrum from physics at extremely high (possibly
trans-Planckian) energies. Although the discussion is usually not
cast in terms of theoretical uncertainties, it is clear that any
intrusion of unknown extremely-high-energy physics into
predictions at an observable level represents an irreducible
obstacle to making meaningful predictions purely using models of
lower-energy physics. Clearly all the marbles are at stake here.
If we must understand trans-Planckian physics to predict what
inflation implies for CMB fluctuations, then the very {\it
predictability} of inflationary models is lost.

These same issues have previously arisen in a much broader
context, since one can ask why high-energy physics doesn't
similarly pollute other predictions in physics, and so bring
theoretical physics more generally to a halt. How this works is
well understood outside of the inflationary context, where it is a
well-established property of quantum field theories that
high-energy modes generically {\it decouple} from lower-energy
phenomena. This is not to say that they are completely irrelevant
at lower-energies, just that their low-energy influence is
channelled through a very few parameters. (For example, atomic
physics {\it does} depend on nuclear physics, but typically only
through a few properties like the mass and charge of the nucleus.)

Although trans-Planckian physics may not be described by a quantum
field theory, it is very likely to decouple in the above sense
inasmuch as our ignorance of its nature has not yet proven to be
an obstacle to understanding lower-energy phenomena. Much of the
exploration of string theory in particular presupposes this
decoupling by expressing low-energy string effects in terms of
effective low-energy quantum field theories. The burden is on any
theory of trans-Planckia which does {\it not} decouple to explain
why low-energy predictability is possible at all.

On the other hand, for inflation ref.~\cite{shenker} argues that
such general decoupling arguments require the influence of physics
at scale $M \gg H$ to contribute at most of order $H^2/M^2$ to
observable effects in the CMB, where $H$ is the Hubble scale at
horizon exit. If true, this would pretty much preclude the
intrusion of all higher-energy physics into CMB fluctuations and
any deviations from the predictions of low-energy models would
necessarily imply the existence of new light degrees of freedom
directly at the epoch of horizon exit. We argue here that although
this conclusion is correct for most effective interactions, there
are loopholes about whose existence one should be aware.

What is it about the models of refs.~\cite{tp1,tp2,tp3} which
allows them to produce observable effects, and so apparently to
fly in the face of decoupling arguments? For some of the models
\cite{tp3}, the observable effects are tied up with the use of the
$\alpha$-vacua of de Sitter space, the validity of which is still
subject to some controversy \cite{dSinconsistent}. For the others,
however, the $\alpha$-vacua are not necessary, with the models
often simply consisting of free particles --- albeit with
unconventional, nonrelativistic dispersion relations --- which
generically {\it should} decouple.

Here we summarize the results of refs.~\cite{TPI1,TPI2}, which
identify the loopholes of the general decoupling arguments on
which the models rely. This is done by studying the implications
of decoupling for inflation using garden-variety hybrid-inflation
models containing high-energy (but {\it sub}-Planckian) physics,
whose low-energy effects may be described using standard methods.
The purpose of using such conservative models is to cleanly
identify what it takes for high-energy effects to appear in the
CMB, and in particular to divorce these inflationary effects from
the more exotic properties of the previously-proposed models (such
as the failure of Lorentz invariance), which are irrelevant and
are likely to be constrained by other, non-inflationary,
measurements.

This study identifies three ways in which such models can alter
inflationary predictions for the CMB, and we believe all of the
non-$\alpha$-vacuum models proposed employ one of these ways. The
three ways apply more generally than just within an inflationary
context, and are:
\begin{enumerate}
\item The model could introduce new degrees of freedom which are
actually not heavy compared with $H$ at the epoch of horizon exit.
This is not really a loophole to decoupling arguments because the
new physics is not heavy and so need not decouple. We include it
in this list for completeness since it is the standard way to
alter the CMB in inflation, and is the way in which most
inflationary models differ from one another. It is also the
mechanism used in some models of \cite{tp1,tp2}, wherein particles
exist having dispersion relations which predict states having very
small energies at large momenta.
\item The model may have rapid time dependence, in the sense that
the states of the low-energy theory do not evolve adiabatically.
There are two common ways for adiabatic evolution to fail. First,
it might happen that the time-dependence of background fields
causes an initially large energy gap between high- and low-energy
states to become small, and so no longer to suppress the amplitude
for exciting the (no-longer-so) `heavy' states. If so the model
becomes an example of the previous case, option 1 above.
Alternatively, the background time-dependence can be rapid enough
to induce direct transitions between what were nominally low- and
high-energy states. In this second case heavy states having
energies which differ from light states by the frequency of the
driving fields typically don't decouple since they may be directly
produced from initial states which only involve the light
particles.
\item Finally, even if the only new physics is heavy and all time
evolution is adiabatic, the low-energy theory can involve relevant
or marginal effective interactions which are sensitive to some of
the details of the high-energy world. Although the implications of
high-energy physics at scale $M$ are {\it generically} smaller
than $O(H^2/M^2)$, they need not always be this small. We identify
effective interactions which depend logarithmically on $M$, and
some which contribute to observables at order $m^2/M^2$, with $m
\gg H$. The inflaton potential is often a good place to look for
such interactions, since they need only compete there with the
very small low-energy inflaton interactions which are consistent
with the very shallow potential which inflation requires.
\end{enumerate}

Although the thrust of these examples is that high-energy physics
{\it can} in some circumstances intrude into CMB fluctuations,
decoupling implies that it does not do so in an uncontrolled way,
and so it does {\it not} introduce uncontrollable theoretical
errors into the predictions of low-energy inflationary models. In
this sense our results represent in some ways the best of all
possible worlds, inasmuch as the broad implications of inflation
are not undermined, but it is also not crazy to look for
deviations from low-energy models in observations.

In what follows we do not further explore option 1, but instead
summarize the examples of options 2 and 3 which are respectively
described in refs.~\cite{TPI1} and \cite{TPI2}.

\section{Non-Adiabatic Physics}
We first describe a simple hybrid-inflation model \cite{hybrid}
for which fluctuations in the CMB bears the imprint of a period of
non-adiabatic oscillations of heavy scalar fields prior to the
epoch of horizon exit. The upshot in this model is that the CMB
can be sensitive to such a non-adiabatic period, but only if they
occur up to 10 $e$-foldings before horizon exit. (Related models
can be constructed for which the CMB can see the implications of
adiabatic physics for up to 30 $e$-foldings before horizon exit
\cite{TPI1}).

We find these kinds of non-adiabatic oscillations generically have
two kinds of implications for the CMB spectrum:
\begin{itemize}
\item They can suppress power in the lowest few multipoles (similar
to what is also seen if a pre-inflationary phase were to end just
before horizon exit), and
\item They can introduce features in the spectrum at specific
wave-numbers which are related to the oscillation frequency of the
non-adiabatically oscillating fields.
\end{itemize}
It is intriguing that there is (currently quite weak) evidence for
both of these predictions in the observed CMB fluctuations, and
the suppression of power on large scales in particular typically
makes the predictions of the models we discuss slightly better
fits to the CMB spectrum as measured by the WMAP collaboration
\cite{Spergeletal} than is the standard concordance cosmology,
although only at roughly the 1.5 sigma level (similar to the
predictions of a pre-inflationary phase \cite{jim}).

\subsection{The Model} \label{S:TheModel}
The lagrangian density we study for these purposes is
\beqa \label{eq:modeldef}
    - {\cal L} &=& \sqrt{-g} \Bigl[\sfrac12  \partial_\mu  \phi \,
    \partial^\mu \phi + \sfrac12 \partial_\mu  \chi \, \partial^\mu \chi +
    V(\phi,\chi) \Bigr] , \\
    \hbox{with} \qquad V(\phi,\chi) &=& \sfrac12 \, m^2 \, \phi^2
    + \, \lambda (\chi^2 - v^2)^2 + \sfrac12 \,
    g \, \chi^2 \phi^2 + \tilde\lambda \, \phi^4. \nonumber
\eeqa
The potential has absolute minima at $\chi = \pm v$ and $\phi =
0$, but also has a long trough at $\chi = 0$ provided $g \, \phi^2
> \lambda \, v^2$. Inflation occurs in the model if the inflaton,
$\phi$, starts at $\phi = \phi_0$ deep along the bottom of the
trough, with $g \phi_0^2 \gg \lambda\, v^2$, and then rolls to
smaller $\phi$ until either the slow-roll parameters become large,
or $g \phi^2 \sim \lambda v^2$, where the $\chi = 0$ minimum is
destabilized. The roll of $\phi$ along the trough bottom can be
sufficiently slow to give inflation provided the various
parameters of the scalar potential are assumed to take appropriate
values.

Under these conditions the field $\chi$ is a heavy degree of
freedom throughout all but the very end of the inflationary epoch,
since its mass is
\beq \label{eq:Mdef}
    M^2 =  - \lambda\, v^2 + g \, \phi^2 \approx g
    \phi^2,
\eeq
which typically satisfies $m \ll H \ll M$ during inflation due to
the assumptions which the inflaton potential must satisfy in order
to produce inflation. We assume $\phi_0$ to be small enough to
ensure $M \ll M_p$ throughout inflation, as is required for us to
maintain theoretical control over all calculations.

\subsection{$\chi$ Oscillations}
The picture so far is standard. Our only modification is to choose
$\chi$ initially not to lie precisely along the bottom of the
trough. Instead we choose the initial values $\chi_0 \ne 0$ and
$\dot{\chi}_0=0$. For simplicity we consider only the evolution of
the homogeneous $\chi$ mode, since this already suffices to show
that nontrivial implications for the CMB are possible. With
$\chi_0$ chosen close enough to the trough's bottom we may neglect
the effect of the $\chi^4$ terms in the potential, leading to
\beq \label{eq:chieqn}
    \ddot\chi + 3 H \, \dot \chi + M^2(\phi) \, \chi \approx 0.
\eeq
The general solution to this equation is known if the initial
condition satisfies $\dot\chi_0/\chi_0 \lsim O(H)$, regardless of
the time-dependence of $H$, so long as we may also neglect
$\dot\phi/\phi$ and $H$ in comparison with $M$. It is
\beq \label{eq:chiform}
    \chi(t) \approx A(t) \cos\Bigl[M(\phi) (t- t_0) \Bigr],
\eeq
where the slowly-varying envelope is given by $A(t) = \chi_0 \,
[a(t_0)/a(t)]^{3/2}$, and $M^2(\phi)\simeq g \phi^2$. By virtue of
the condition $M \gg H$ discussed above, this evolution describes
a fast oscillation rather than a slow roll relative to the
timescale set by the expansion of the universe.

The energy density associated with these oscillations is
\beq \label{eq:chiEnergy}
    \rho_\chi(t) = \sfrac12 \Bigl( \dot\chi^2 + M^2 \, \chi^2
    \Bigr) = \sfrac12 \, M^2 \, \chi_0^2 \,
    \left( {a(t_0) \over a(t)} \right)^3 ,
\eeq
which scales with $a(t)$ as does non-relativistic matter. The
amplitude of the $\chi$ oscillations is damped by the Hubble
expansion, and so long as the $\phi$ roll remains slow an
inflationary phase eventually begins. Whether inflation occurs
depends on how far the inflaton has rolled down its trough in the
time taken for the $\chi$ oscillations to be damped away. We
assume the initial conditions $\phi_0$ and $\chi_0$ to be chosen
to ensure that sufficient inflation does occur after the $\chi$
oscillations become negligible.

It is convenient to choose our initial time, $t_0$, as the time
when the energy of the $\chi$ oscillations first becomes small
enough to allow inflation to begin. In this case the amplitude
$\chi_0$ may be found by equating the $\chi$-oscillation energy,
$\sfrac12 \, M^2 \, \chi_0^2$ to the inflationary vacuum energy,
$\sfrac14 \, \lambda \, v^4 = 3 H^2 M_p^2$, implying
\beq
    \chi_0^2 = {6 H^2 M_p^2 \over M^2} \, .
\eeq
Because the $\chi$ oscillations are damped during inflation
proportional to $[a_0/a(t)]^3 = \exp[-3H(t - t_0)]$, we see that
the number of $e$-foldings between the beginning of inflation and
horizon exit is related to the oscillation amplitude, $\chi_{he}$,
at horizon exit by
\beq \label{eq:efoldsize}
    H \, \tau \sim \sfrac13 \, \ln \left( {\chi_0^2 \over \chi_{he}^2}
    \right) \sim \sfrac13 \, \ln \left( { 6 M_p^2/M^2
    \over \chi_{he}^2/H^2} \right) ,
\eeq
where $\tau = |t_{he}-t_0|$. Clearly --- for fixed fluctuation
size, $\chi_{he}/H$ --- the later horizon exit occurs after the
onset of inflation, the lower $M$ must be, and hence the smaller
$m$, $v$ and $H$ must also be in order to have sufficient
inflation {\em after} horizon exit. This last formula is useful
because it is convenient to use $\chi_{he}$ to parameterize the
amplitude of oscillations, since this parameter directly controls
the size of the oscillation effects seen in the CMB.

\subsection{Modifications to Inflaton Fluctuations}
\label{S:Oscillations}
We now ask how the $\chi$ oscillations change the power spectrum
of inflaton fluctuations which get imprinted onto the CMB. To this
end consider the quantum fluctuations of the inflaton, $\dphi =
\phi - \langle\phi\rangle$, having wave number $k$. (Notice
$\langle \phi_k \rangle = 0$ for $k \ne 0$ by virtue of our
assumption that $\chi$ experiences a homogeneous, $k$-independent
roll.) To linearized order in the fluctuations this satisfies the
equation of motion
\begin{equation}\label{eq:fluc}
    \ddot\varphi_k + 3H\dot\varphi_k + \left[ k^2e^{-2Ht} +V''(
    \langle\phi\rangle)
-g \chi^2(t)\right]\dphi_k = 0.
\end{equation}

The $\chi$ oscillations affect the mode functions $\varphi_k$
largely by introducing a time dependence to the `mass' term, $V''
- g \chi^2$, both through its explicit $\chi$-dependence and
through the change the $\chi$ oscillations induce in the evolution
of the background field, $\langle \varphi(t) \rangle$. Numerically
computing the evolution of the background fields, and using these
to solve eq.~\pref{eq:fluc} for the mode which agrees at $t = t_0$
with the usual positive-frequency (Bunch-Davies) mode,
$\varphi_k^+$, leads to a solution $\tilde\varphi_k^+ =
\varphi_k^+ + \delta\varphi_k^+$ whose features are shown in fig.\
\ref{fig:raw}. Notice that the deviation of $|\tilde\varphi_k ^{+}
(\infty)|$ from the value $|\varphi_k^+(\infty)| = H$ it would
have had in the absence of $\chi$ oscillations can be large, even
if $\chi_{he}$ is small.

The fractional deviation in the power spectrum is then computed
using $\delta P_k/P_k = |\tilde\varphi_k^+/\varphi_k^+|^2 -1$,
with the right-hand-side evaluated as $t \to \infty$. In figure
\pref{fig:kdiff} we plot the log of the absolute value of the
percentage deviation ($\log_{10}(|\delta P|/ P \times 100)$) as a
function of $\log_{10}(k/H)$, for a range of values of $M$, and
for two different values of $t_0$.

\FIGURE{ \epsfig{file=raw-pap.eps, width=3.5in}
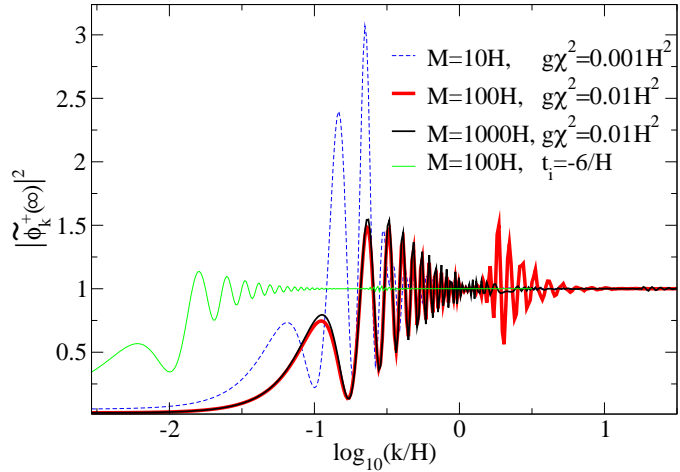
        \caption[Figure 1]{$|\tilde\varphi_k ^{+}(\infty)|^2$ in units
        of $H^2$ for the hybrid inflation
        model, as a function of $\log_{10}(k/H)$
        for several values of $M$ and $g\chi_{\rm he}^2$, rightmost
        three curves for $t_0 = -4/H$.
Leftmost curve shows the effect of taking an earlier initial time,
$t_0=-6/H$, with $g\chi_{\rm he}^2=10^{-6}$. }\label{fig:raw}}

\FIGURE{ \epsfig{file=kdiff-pap.eps, width=3.5in}
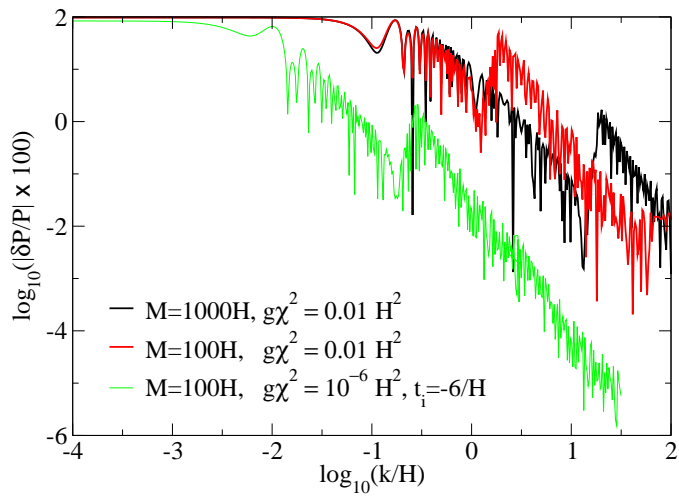
        \caption[Figure 2]{Log of absolute value of percent deviation of power
spectrum as a function of $\log_{10}(k/H)$, for $M=100H$ and
$M=1000H$, with $t_0= -4/H$ and $g\chi_{\rm he}^2=0.01 H^2$, and
$M=100H$ with $t_0=-6/H$ and $g\chi_{\rm he}^2=10^{-6} H^2$.
Notice that the deviation is large at low $k$ not because the
power is large, but rather because it is smaller than normal.
Order of curves in legends coincides with that at right hand edge
of the graph. }\label{fig:kdiff}}

These figures show three main features, each of which has a simple
physical explanation \cite{TPI1}.
\begin{enumerate}
\item The fluctuation spectrum oscillates rapidly, due to the
rapid driving by the fast $\chi$ oscillations.
\item The envelope of the oscillating inflaton fluctuations is
strongly suppressed for the lowest $k$ values. This is because at
early times the average value of $g\chi^2$ is potentially large,
since $\langle \cos^2(Mt) \rangle = \sfrac12$, and so the
oscillating $\chi$ field behaves like an inflaton mass, and like a
mass it suppresses fluctuations for small $k$. (It is because the
CMB observations also appear to point to a similar suppression for
small $k$, that these models provide a slight improvement in
goodness-of-fit over fits to the standard concordance cosmology.
The comparison of these models with the WMAP data is very similar
to that given in ref.~\cite{jim}.)
\item The fluctuation spectrum rises to a peak, whose position is
easily understood in terms of the driving frequency of the $\chi$
oscillations. These oscillations resonantly excite $\varphi$ modes
having the same frequency, but the wavelength of these modes
redshift as the universe expands. The peak occurs for wave-numbers
corresponding to those modes which were driven at the earliest
times, $t = t_0$, since this was the point when the driving $\chi$
field had the largest amplitude.
\end{enumerate}

Detailed comparisons with the spectrum of CMB fluctuations show
that an amplitude at horizon exit, $g\chi_{he}^2=10^{-5} H^2$
corresponds to a roughly 5\% change in the CMB spectral
parameters, making this a rough benchmark for how large an
oscillation must be in order to have detectable effects. Given
this benchmark, eq.~\pref{eq:efoldsize} tells us how long before
horizon exit inflation can have lasted without being overwhelmed
by the energy in the damped $\chi$ oscillations. Pushing all
parameters to make this time as long as possible leads in this
model to $H\tau \lsim 10$ $e$-foldings. A similar limit for $\tau$
is also obtained if one asks that the $\chi$ oscillations not lose
all of their energy through decays into inflaton quanta. In other,
less well motivated, models Ref.~\cite{TPI1} a similar analysis
shows heavy-field oscillations can have observable consequences
for up to 30 $e$-foldings before horizon exit.

In summary, this model produces observable implications for the
CMB because the fast motion of the background $\chi$-field makes
the inflaton evolution non-adiabatic, and therefore causes
positive- and negative-frequency modes to mix. Even outside of an
inflationary context, nobody would expect to be able to describe
the low-energy inflaton physics in the presence of these $\chi$
oscillations using an effective theory within which the $\chi$
field had been integrated out.

\section{Adiabatic Physics}
In this section we consider a model similar to the one examined
above, but in which we make the more standard assumption that the
background heavy fields do not oscillate. As a result the time
evolution of the inflaton field is adiabatic, and the influence of
the heavy fields is well described by a low-energy effective
theory involving only the light degrees of freedom. For technical
reasons we couch this section's discussion in terms of a
supersymmetric extension of the model just discussed. We use a
supersymmetric model so that the heavy-field effective
contributions to the inflaton potential do not destroy its
flatness.

As is often the case with supersymmetric theories, in the model we
consider the tree-level inflaton potential is exactly flat, but
this flat direction is lifted by virtual loops of heavy particles.
We show that the potential depends logarithmically on the heavy
mass, leading to slow-roll parameters which are suppressed by
factors of order $M_0^2/M^2$, where $M$ is the heavy mass but
$M_0$ can be much larger than $H$. The model shows that heavy
physics can decouple and yet still alter inflationary predictions
for the CMB, since the figure of merit for deciding the
observability of the heavy-physics effects can be larger than
$H^2/M^2$.

\subsection{The Model}
Consider a globally-supersymmetric model containing the chiral
multiplets, $\Phi = \{\phi,\psi\}$ and $H_\pm =
\{h_\pm,\chi_\pm\}$, coupled to a $U(1)$ gauge multiplet, $V =
\{A_\mu,\lambda\}$. $\Hp$ and $\Hm$ carry opposite $U(1)$ charges
$\pm e$, and the multiplet $\Phi$ is neutral. The model's
superpotential and K\"ahler potential are
\beq
    K = \Hp^* \Hp + \Hm^* \Hm + \Phi^* \Phi \qquad \hbox{and}
    \qquad W = g \Phi \, (\Hp \, \Hm - v^2) \, ,
\eeq
where $g$ and $v$ are real constants. The associated scalar
potential for this theory is $V = V_F + V_D$ where
\beqa
    V_F &=& g^2 \left( \Bigl| \hp \hm - v^2 \Bigr|^2 + \Bigl|  \phi \,
    \hm \Bigr|^2 + \Bigl|  \phi \,  \hp \Bigr|^2 \right) \, , \nonumber \\
    V_D &=& \frac{e^2}{2} \, \Bigl( |\hp|^2 - |\hm|^2 + \xi
    \Bigr)^2 \, ,
\eeqa
and $\xi > 0$ is the Fayet-Iliopoulos term. The global minimum is
supersymmetric with
\beq
    \phi = 0, \qquad |h_\pm|^2 = \frac12 \Bigl( \mp \xi + \sqrt{\xi^2
    + 4 v^4} \Bigr) \, ,
\eeq
at which point $V = 0$. There is a trough at $h_\pm = 0$, for
large $|\phi|$, along which $V_{\rm trough}(\phi) = V(h_\pm = 0,
\phi) = g^2 \, v^4 + \sfrac12 \, e^2 \xi^2$ is independent of
$\phi$. The potential's curvature in the $h_\pm$ directions is
\beq
    M_\pm^2(\phi) = g^2 |\phi|^2 \pm \Delta \,
    ,
\eeq
with $\Delta = \sqrt{ g^4 v^4 + e^4 \xi^2}$, showing that the
$h_\pm$ masses are positive for all $|\phi|^2 > \Delta/g^2$, with
masses which get bigger the larger $|\phi|$ is.

Along the trough's bottom the gauge bosons are massless since the
$U(1)$ gauge invariance is unbroken. The fermions $\lambda$ and
$\psi$ are massless at tree level, while the fermions $\chi_\pm$
have masses
\beq
    m^2_\pm(\phi) = m^2(\phi) = g^2 |\phi|^2 \, .
\eeq
We therefore find a low-energy sector of strictly massless
particles, $\{A_\mu,\lambda,\phi,\psi\}$, which do not classically
directly couple among themselves but which do couple to a massive
sector, $\{\hp,\chi_+,\hm,\chi_-\}$. Our interest is in the
effective interactions which are generated amongst the light
fields once these heavy modes are integrated out.

Integrating out the heavy fields leads to the following one-loop
contribution to the low-energy scalar potential,
\beqa \label{eq:effpotgs}
    V_{\rm eff}(\phi) &=& \rho + \Delta V(\phi), \\
    \hbox{with} \qquad \Delta V(\phi) &=& \delta \rho +\frac{2N}{64
    \pi^2} \sum_{i=\pm} \left[ M_i^4(\phi) \, \ln \left( {M_i^2(\phi)
    \over \mu^2} \right) - m_i^4(\phi) \, \ln \left( {m_i^2(\phi)
    \over \mu^2} \right) \right] \, ,
    \nonumber
\eeqa
where $\rho = g^2 \, v^4 + \sfrac12 \, e^2 \xi^2$ is the
renormalized (constant) classical potential along the trough, and
$\delta \rho$ is the corresponding counter-term. (The overall
factor of $N$ arises if we extend the model to include $N$ heavy
multiplets, all sharing the same tree-level couplings to the light
fields.) For $m^2(\phi) \gg \Delta$ this becomes
\beq
    \Delta V_{\rm eff}(\phi) \approx  \frac{N \, \Delta^2}{16
    \pi^2}\left[ \ln \left( {m^2(\phi) \over m_*^2} \right)
    + {\cal O} \left( {\Delta^2 \over m^4 } \right) \right] \, ,
\eeq
where we adopt the renormalization condition that $\Delta V$ must
vanish when $\phi = \phi_*$, defined as the field's value at
horizon exit. If $N_e$ $e$-foldings of inflation occur between
horizon exit and the end of inflation, we have
\beq
    m_*^2 = m^2(\phi_*) \approx m^2_{\rm end} + \frac{g^2 N N_e
    \Delta^2}{12 \pi^2 H^2} \approx m^2_{\rm end} +
    \frac{g^4 N N_e \, M_p^2}{4 \pi^2} ,
\eeq
where $m_{\rm end}^2$ is either $\Delta$ or $g^4 N M_p^2/(8
\pi^2)$, whichever is larger. These estimates use the natural
choice $e \sim g$ and $\xi \sim v^2$, for which $\Delta \sim g^2
v^2$ and $\rho \sim g^2 v^4$.

At horizon exit the inflationary parameters \cite{liddlelyth}
which are predicted by this potential are
\beqa
    H^2 &=& \frac{V}{3 M_p^2} \approx \frac{\rho}{3 M_p^2} \approx
    \frac{g^2 v^4}{3 M_p^2} \, , \nonumber \\
    \epsilon_* &=& \frac12 \left[ {M_p \over V_{\rm
    eff}} \left( {\partial V_{\rm eff} \over
    \partial \varphi} \right)\right]^2_{\varphi_*}  \approx \frac12 \left[ {M_p \,N \,
    \Delta^2 \over 8 \pi^2 \rho \, \varphi_*} \right]^2
    \approx \left( \frac{g^2 N }{32\pi^2 N_e} \right) \, , \nonumber \\
    \eta_* &=& {M_p^2 \over V_{\rm eff}} \left( {\partial^2
    V_{\rm eff} \over \partial \varphi^2} \right)_{\varphi_*}
    \approx - \, {M_p^2 \,N\, \Delta^2 \over 8 \pi^2
    \rho\, \varphi^2_*}
    \approx -\, \frac{1}{2 N_e}
    \, ,
\eeqa
where $\varphi = |\phi|$ is the inflaton. These equations shows
that the roll is sufficiently slow if $\varphi/M_p \gsim g/4\pi$.
Notice that if $\varphi$ should be as large as $O(M_p)$
consistency would require us to embed this model into
supergravity, a situation we can avoid if $g \ll 1$. For $N=1$ all
of the requirements for inflation with $N_e \gsim 60$ are
satisfied --- including fluctuation amplitudes which agree with
CMB observations --- if $v/M_p \sim g \sim 10^{-3}$.

For the purposes of comparing to decoupling arguments notice that
the slow-roll parameters may be written
\beq
    (2 \epsilon_*)^{1/2} \approx \frac{g^2 N \Delta^2}{8 \pi^2 \, \rho} \, \left(
    \frac{\varphi_* \, M_p}{m^2_*} \right) \qquad \hbox{and}
    \qquad \eta_*  \approx -\,\frac{g^2 N \Delta^2}{8 \pi^2 \,
    \rho} \, \left(\frac{M_p^2}{m^2_*} \right) \,,
\eeq
which implies the heavy physics decouples, inasmuch as both $\eta$
and $\epsilon$ are suppressed by inverse powers of the heavy mass,
$m(\phi)$. But the scale against which $m^2(\phi)$ is compared is
not $H^2$, but is instead either $(g^4 N/8\pi^2) M_p \, \varphi$
or $(g^4 N/8 \pi^2) M_p^2$. (The condition $m(\phi) \ll M_p$
clearly again implies the coupling $g$ must be small.) If
parameters are adjusted so that the heavy physics scale, $m_*$, is
dialed to become larger and larger with $H$ fixed, then the slow
roll parameters decrease, becoming closer and closer to the
scale-invariant prediction $\epsilon_* = \eta_* = 0$. This
expresses the consequences of decoupling, since the entire
inflaton potential is generated by virtual effects of the heavy
physics. But because the benchmark for observability in this case
is {\it not} $H^2/m^2_*$, the difference in their predictions can
be kept observable even if $H^2/m_*^2$ is much smaller than a few
percent.

In particular, imagine now comparing the effects for the CMB of
two theories which differ only in that one has $N = 1$ and the
other has $N=2$ heavy sectors. If we suppose both models to
undergo the same number of $e$-foldings of inflation, then they
must also agree on their predictions for $\eta$. They can also
predict identical fluctuation amplitudes so long as $v_1^4 =
v_2^4/2$, since $\delta\rho/\rho \propto H^2/(M_p^2 \epsilon_*)
\propto (v/M_p)^4 (N_e/N)$. If they share the same couplings, $e
\sim g$, then the models will predict $\epsilon_{*2} = 2
\epsilon_{*1}$, and so can have detectable differences in their
predictions for CMB observables.

\acknowledgments

C.B. would like to thank the organizers for their invitation to
speak. We thank Robert Brandenberger, Brian Greene, Nemanja
Kaloper, Anupam Mazumdar and Steve Shenker for helpful
discussions. R.~H. was supported in part by DOE grant
DE-FG03-91-ER40682, while the research of C.B., J.C. and F.L. is
partially supported by grants from McGill University, N.S.E.R.C.
(Canada) and F.C.A.R. (Qu\'ebec).

\end{document}